\documentclass[aps, twocolumn, letterpaper]{revtex4}

\pdfoutput=1

\usepackage{amsmath}
\usepackage{amssymb}
\usepackage{graphicx}
\usepackage{xspace}
\usepackage{upgreek}
\usepackage{accents}

\newcommand{\eg}{{e.g.,\/}\xspace}
\newcommand{\ie}{{i.e.,\/}\xspace}

\newcommand{\eq}[1]{(\ref{#1})}
\newcommand{\Eq}[1]{Eq.~(\ref{#1})}
\newcommand{\Eqs}[1]{Eqs.~(\ref{#1})} 
\newcommand{\Fig}[1]{Fig.~\ref{#1}} 
\newcommand{\Ref}[1]{Ref.~\cite{#1}}

\newcommand{\Sec}[1]{Sec.~\ref{#1}}

\newcommand{\mc}[1]{\mathcal{#1}}

\newcommand{\favr}[1]{\langle #1 \rangle}
\renewcommand{\vec}[1]{{\boldsymbol{\rm #1}}}

\newcommand{\pd}{\partial}

\newcommand{\del}{\nabla}
\newcommand{\kpt}[1]{{\kern #1 pt}}
\newcommand{\const}{\text{const}}

\newcommand{\msection}[1]{\textit{#1.}\ ---\ }  

\begin{document}
\title{Surfatron acceleration along magnetic field by oblique electrostatic waves}
\author{I.~Y. Dodin and N.~J. Fisch}
\affiliation{Department of Astrophysical Sciences, Princeton University, Princeton, New Jersey 08544, USA}
\noaffiliation
\affiliation{Princeton Plasma Physics Laboratory, Princeton, New Jersey 08543, USA}
\noaffiliation

\begin{abstract}
Charged particles can be accelerated via surfatron mechanism along dc magnetic field by obliquely propagating electrostatic waves. In plasma, this mechanism can, in principle, produce an average parallel current, even when the wave frequency is much larger than the gyrofrequency and the wave phase velocity is much larger than particle initial velocities.
\end{abstract}

\maketitle

\section{Introduction}

It is well known that a charged particle can be accelerated resonantly by a wave along the phase front in the presence of weak dc magnetic field $\vec{B}_0$ \cite{foot:review}. This so-called surfatron acceleration was described in \Ref{ref:dawson83} for $\vec{B}_0$ perpendicular to the wave vector $\vec{k}$. In particular, it was shown that speeds of the order of the $\vec{E} \times \vec{B}_0$ velocity can be attained for nonrelativistic particles, $\vec{E}$ being the wave electric field. A more general case, with the angle between $\vec{B}_0$ and $\vec{k}$ being arbitrary, is more complicated, and it is only recently that the underlying physics was made transparent, namely, via Hamiltonian geometrical arguments in \Ref{ref:vasiliev11}. However, those arguments do not immediately apply to electrostatic waves, which also could drive surfatron acceleration in plasma (\eg ion acceleration in the lower-hybrid frequency range). Hence, an operational understanding of the surfatron effect in oblique electrostatic waves is yet to be developed.

Here, we study this effect in application to nonrelativistic particles. We proceed like in \Ref{ref:vasiliev11}, but without using the Hamiltonian formalism explicitly, striving to render the qualitative picture even more transparent. For the particular case when $\vec{k}$ is perpendicular to $\vec{B}_0$, the corresponding result of \Ref{ref:dawson83} is recovered. Otherwise, acceleration \textit{along} magnetic field is possible, like in \Ref{ref:vasiliev11}; yet, the particle dynamics in an electrostatic wave can be qualitatively different from that in an electromagnetic wave (\Sec{sec:single}). We also find that, in plasma, surfatron mechanism can, in principle, produce an average parallel current, even when the wave frequency is much larger than the gyrofrequency and the wave phase velocity is much larger than particle initial velocities.

The presentation is organized as follows. In \Sec{sec:basic}, we introduce basic equations. In \Sec{sec:single}, we identify and explain the possible types of particle trajectories. In \Sec{sec:cd}, we discuss current drive via surfatron acceleration. In \Sec{sec:conc}, we summarize our main results.

\section{Basic equations}
\label{sec:basic}

\begin{figure*}
\centering
\includegraphics[width=.99\textwidth]{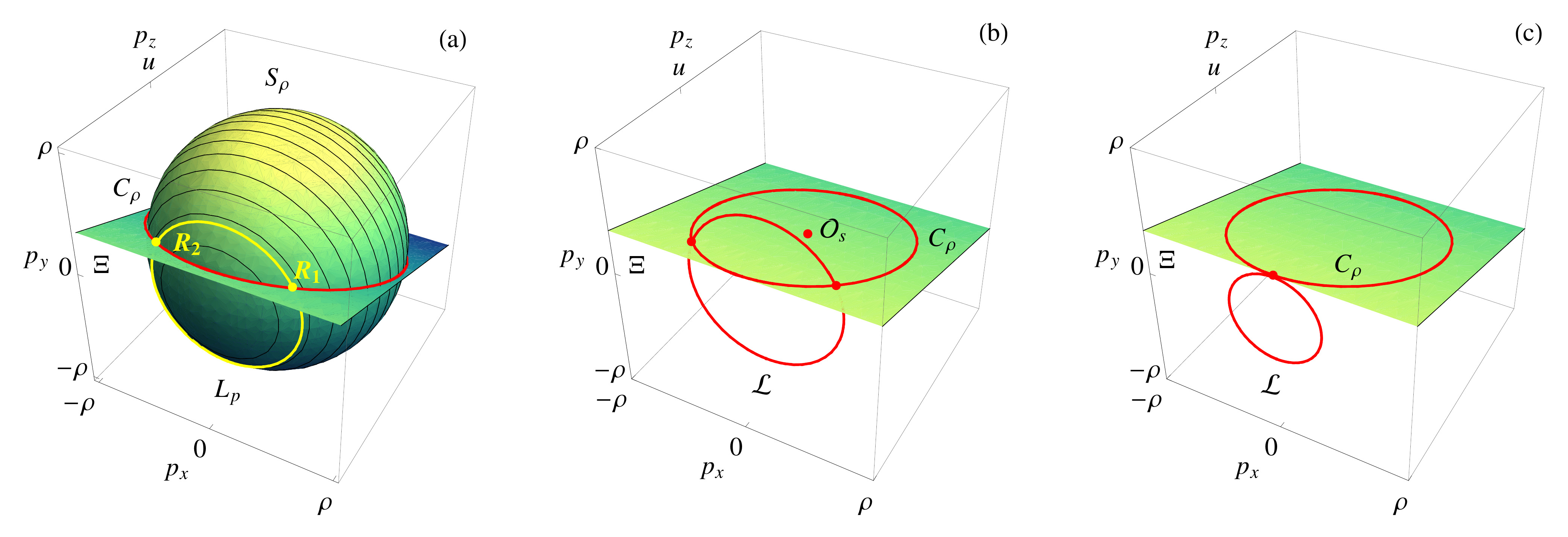}
\caption{(Color online) Momentum space geometry. (a)~Here $S_\rho$ is the sphere defined by \Eq{eq:S}; $\Xi$ is the resonance plane defined by \Eq{eq:Xi}; $C_\rho$ is the circle obtained as the intersection of $S_\rho$ and $\Xi$; $L_p$ is a Larmor orbit; $R_{1,2}$ are the points where $L_p$ intersects with $\Xi$ (and $C_\rho$ also). (b)~Here $\mc{O}_s$ is the center of $C_\rho$ (and of $S_\rho$ also); the ring $\mc{L}$ is as defined in \Sec{sec:cd}. (c)~The ring $\mc{L}$ with the radius $p_{\perp 0} = \bar{p}(p_{z0})$, in which case $\Xi$ is tangent to $C_\rho$.}
\label{fig:geom}
\end{figure*}

Suppose a dc magnetic field of the form $\vec{B}_0 = \vec{z}^0 B_0$ and a wave field $\vec{E} = -\del \phi$, with some frequency $\omega$ and the wavevector of the form $\vec{k} = \vec{y}^0 k_y + \vec{z}^0 k_z$; namely,
\begin{gather}
\vec{\phi} = a \cos \psi, \quad \psi = \omega t - k_y y - k_z z.
\end{gather}
(For clarity, we do not constrain the wave to obey a specific dispersion relation.) Then, the equation for the particle momentum $\vec{p}$ reads~as
\begin{gather}
\dot{\vec{p}} = e \del \phi - \Omega\,\vec{p} \times \vec{z}^0,
\end{gather}
where $\Omega = eB_0/(mc)$ is the gyrofrequency, $- e$ and $m$ are the particle charge and mass, and $c$ is the speed of light. Let us measure $\vec{p}$ in units $mc$, $a$ in units $mc^2/e$, $\Omega$ in units $\omega$, $\vec{k}$ in units $\omega/c$, spatial coordinates in units $c/\omega$, and time $t$ in units $\omega^{-1}$. Then, 
\begin{align}
& \dot{p}_x = - p_y \Omega, \label{eq:px}\\
& \dot{p}_y = p_x \Omega + k_y a \sin \psi, \label{eq:py}\\
& \dot{p}_z = k_z a \sin \psi, \label{eq:pz}\\
& \dot{\psi} = 1 - k_y p_y - k_z p_z, \label{eq:dpsi}
\end{align}
or, alternatively,
\begin{gather}
\ddot{p}_x + \Omega^2 p_x = - \Omega k_y a \sin \psi, \label{eq:ddpx}\\
\ddot{\psi} + k^2 a \sin \psi = - \Omega k_y p_x. \label{eq:ddpsi}
\end{gather}
Notice that \Eqs{eq:px}-\eq{eq:ddpsi} have an integral $p_x^2 + p_y^2 + (p_z - k_z^{-1})^2 - 2 a \cos \psi = \const$, owing to the energy conservation in the frame where the wave is stationary \cite{foot:frame}. Assuming that the wave is weak, the particle motion is thus approximately bound to a sphere $S_\rho$ in $\vec{p}$-space [\Fig{fig:geom}(a)],
\begin{gather}\label{eq:S}
p_x^2 + p_y^2 + (p_z - u)^2 = \rho^2,
\end{gather}
with the center being at $\mc{O}_s \equiv (0,0,u)$. Here $\rho$ the radius determined by initial conditions, $u = (\cos \alpha)^{-1}$ is the wave longitudinal phase velocity, and $\alpha$ is defined as
\begin{gather}
\sin \alpha = k_y/k, \quad \cos \alpha = k_z/k.
\end{gather}
Below, we will assume, for clarity, that $0 < \alpha < \pi/2$, so, in particular, $u > 0$.

\section{Single-particle motion}
\label{sec:single}

For simplicity suppose $k_y \sim k_z$; then, it is easy to see that the characteristic frequencies in \Eqs{eq:ddpx} and \eq{eq:ddpsi} are $\Omega$ and $\omega_E \equiv k\sqrt{a}$. Of interest for us will be the essentially nonlinear regime, when
\begin{gather}\label{eq:Theta}
\Theta \equiv \Omega/\omega_E \ll 1.
\end{gather}
[Notice that, at $a \ll p^2$ and $k p \lesssim 1$, one also has $\omega_E \ll 1$, so \Eq{eq:Theta} automatically implies $\Omega \ll 1$ as well.] In this case, the particle nonlinear dynamics is quasiadiabatic and can be understood as follows.

\subsection{Two-scale oscillations}

First, let us consider the characteristic features of the two types of oscillations, employing \Eqs{eq:ddpx} and \eq{eq:ddpsi}. For the motion with $\pd_t \sim \omega_E$, use that $\ddot{p}_x \gg \Omega^2 p_x$, so one obtains $p_x \lesssim \Omega k_y /k^2$; hence, $p_x^2/a \lesssim {\Theta^2 \ll 1}$. The amplitude of $p_y$-oscillations is larger, namely, $p_y \sim p_x/\Theta$ [\Eq{eq:px}]; yet still $p_y^2/a \sim \Theta^{-2}(p_x^2/a) \lesssim 1$. Also, \Eq{eq:pz} yields $\tilde{p}_z \lesssim k a/\omega_E$ for the quiver part of $p_z$; hence, ${\tilde{p}_z^2/a \lesssim 1}$. Assuming the wave is weak, oscillations at frequencies of the order of $\omega_E$ are thus of small amplitude in $\vec{p}$-space. In contrast, at $\pd_t \sim \Omega$, we have $\ddot{\psi} \ll \omega_E^2 \psi$; then, at $\sin \psi \sim 1$, one gets $p_x \sim p_y \sim p_z \sim V$, where $V \equiv k a/\Omega$ is the characteristic electric-drift velocity \cite{foot:eps}. Since $V^2/a = \Theta^{-2} \gg 1$, oscillations at frequencies of the order of $\Omega$ can be of large amplitude in $\vec{p}$-space. 

Since high-frequency oscillations of $p_x$ are negligible (albeit not so for $p_y$ and $\tilde{p}_z$), \Eq{eq:ddpsi} yields that the dynamics in space $\Psi \equiv (\psi, \dot{\psi})$ is similar to that of a nonlinear pendulum with a bias torque. Hence, two regimes are possible: (i) If $p_x > p_c \equiv k^2a/(\Omega k_y)$, there is no local equilibrium. Thus, all trajectories are untrapped; specifically, $\psi$ increases or decreases indefinitely, resulting in that the right-hand side of \Eq{eq:ddpx} averages to zero on time scales of the particle gyromotion. Then the wave has little effect on the particle dynamics, which thus represents (weakly perturbed) Larmor rotation with frequency $\Omega$ in the plane $(p_x, p_y)$, with $p_z$ approximately fixed. (ii) If $p_x < p_c$, most of trajectories are still untrapped. However, if a particle is close enough to the resonance ($\dot{\psi} \lesssim \omega_E$), it can be trapped near one of the local equilibria, $\psi = 2 \pi n$, where $n$ is integer. [We assume $a > 0$; otherwise, the equilibria would be $\psi = 2\pi (n + 1)$.] Since the parameters of the potential well for $\psi$ change on the time scale large compared to that of the bounce oscillations, the phase area $J = \oint \dot{\psi}\,d\psi$ captured by the trajectory in $\Psi$ is an adiabatic invariant. Hence, the particle remains trapped as long as $J$ stays less than the separatrix value $J_s(p_x(t))$, or, equivalently, while $p_x(t) < p_s(J)$, where $p_s(J) \le p_c$ is the function inverse to $J_s(p_x)$.

\subsection{Types of trajectories}

Since $\psi$ of a trapped particle oscillates around a fixed value, averaging of \Eq{eq:dpsi} over its fast motion yields that such a particle is bound to the ``resonance plane''~$\Xi$ in $\vec{p}$-space [\Fig{fig:geom}(a)], namely,
\begin{gather}\label{eq:Xi}
1 - k_y p_y - k_z p_z = 0,
\end{gather}
which passes through the center $\mc{O}_s$ of each sphere $S_\rho$. The intersection of $S_\rho$ and $\Xi$ is a circle $C_\rho$ with the same center $\mc{O}_s$ and radius $\rho$, yielding that $p_x$ is bounded from above by $\rho$. Hence, if $\rho < p_s(J)$, the particle is trapped forever. If $p_s(J) < \rho < p_c$ or $p_s(J) \le p_c < \rho$, then detrapping occurs at $p_x = p_s(J)$. In particular, for most deeply trapped particles, the latter corresponds to detrapping at $p_x = p_s(J = 0) \equiv p_c$. (This is the surfatron acceleration discussed in \Ref{ref:dawson83}, where zero $J$ and $\alpha$ are taken, also rendering $\rho$ infinite.) Hence, the types of trajectories that can exist in the system are as follows.

\begin{figure*}
\centering
\includegraphics[width=.99\textwidth]{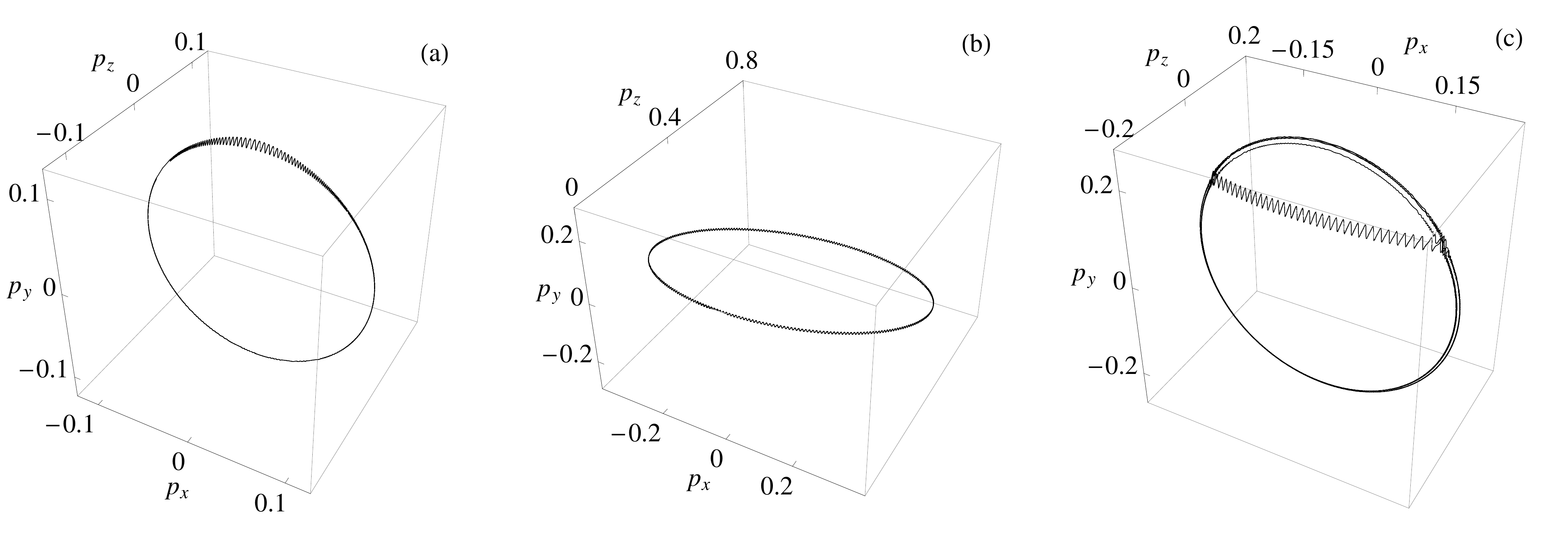}
\caption{Examples of particle trajectories of Type~A, B, and C, respectively, obtained via numerical integration of \Eqs{eq:px}-\eq{eq:dpsi} with $\Theta = 10^{-2}$, $k = 8$, $\alpha = 70^{\circ}$, $a = 5.3 \times 10^{-5}$, $\Omega = 5.8 \times 10^{-4}$, and $u = 0.37$. The initial momentum in Cartesian coordinates is $(p_{\perp 0} \cos \theta,\, p_{\perp 0} \sin \theta,\, p_{z0})$, where $p_{z0} = 0$, so $\bar{p}_\perp(p_{z0}) = k_y^{-1} \approx 0.13$. (a)~$p_{\perp 0} = 0.9 k_y^{-1}$, $\theta = 3\pi/2$, $\psi = \pi/8$; (b)~$p_{\perp 0} = k_y^{-1}$, $\theta = \pi/2$, $\psi = \pi/8$; (c)~$p_{\perp 0} = 2k_y^{-1}$, $\theta = 3\pi/8$, $\psi = 7\pi/8$.}
\label{fig:traj}
\end{figure*} 

\msection{Type A} Particles can be untrapped forever and rotate adiabatically in the plane $(p_x, p_y)$ with frequency $\Omega$; $p_z$ remains approximately fixed then [\Fig{fig:traj}(a)]. For that, the particle Larmor orbit $L_p$ must not intersect $\Xi$; \ie its radius $p_\perp$ must satisfy $p_\perp < \bar{p}_\perp(p_z)$, where
\begin{gather}
\bar{p}_\perp(p_z) \equiv k_y^{-1}|1-k_z p_z|.
\end{gather}

\msection{Type B} Particles can be trapped forever and rotate adiabatically along circles $C_\rho$ in the plane $\Xi$, specifically around the equilibrium at $\mc{O}_s$ and in the direction determined by the sign of $\Omega$ [\Fig{fig:traj}(b)]. The frequency of these oscillations in the limit of small $J$ equals $\Omega \cos \alpha$, as seen from \Eqs{eq:ddpx} and \eq{eq:ddpsi} after linearization for $\psi \ll 1$.

\msection{Type C} Particles can undergo nonadiabatic dynamics which is a mixture of trapped and untrapped motion [\Fig{fig:traj}(c)]. Suppose a particle starts out on a Larmor circle like of Type~A yet with $p_\perp > \bar{p}_\perp(p_z)$. Then there are two intersections of $L_p$ and $\Xi$, at some $p_x = \pm p_r$, where the resonance condition \eq{eq:Xi} is satisfied; we will call those resonant points $R_{1,2}$. There, the particle crosses the separatrix in $\Psi$ and thus has a probability of hopping on the trapped orbit $C_\rho$, if arriving at the appropriate phase; namely, required is [cf. \Eq{eq:ddpsi}]
\begin{gather}\label{eq:approp}
\pm p_r \approx - p_c \sin \psi,
\end{gather}
which also implies that $p_r < p_c$ is needed. Then the particle would have to continue along $C_\rho$ toward smaller $|p_x|$ from, say, $R_1$ and toward larger $|p_x|$ from $R_2$. (Remember that the direction of motion along $C_\rho$ is determined by the sign of $\Omega$ only.) Yet notice that the phase area $J$ that is captured at $R_{1,2}$ equals $J_s(p_r)$ \cite{ref:timofeev78, ref:neishtadt86, ref:cary86}, and thus further increase of $|p_x|$ would result in immediate untrapping. Thus, actual trapping of an initially untrapped particle can occur at $R_1$ but not at $R_2$. 

Hence, the overall corresponding motion consists of the gyromotion (at angular velocity $\Omega$) which is interrupted incidentally at $R_1$; then the particle proceeds to $R_2$ along $C_\rho$ (at angular velocity of about $\Omega \cos \alpha$) and gets untrapped again, returning to the \textit{same} Larmor orbit. Interestingly, for the type of waves considered in \Ref{ref:vasiliev11} realized is a very different scenario. Specifically, only large $p_x$ (rather than small $p_x$, in our notation) are allowed there after trapping, thus resulting in particle jumping between \textit{two} Larmor orbits with the same $p_\perp$.

\begin{figure}[b]
\centering
\includegraphics[width=.3\textwidth]{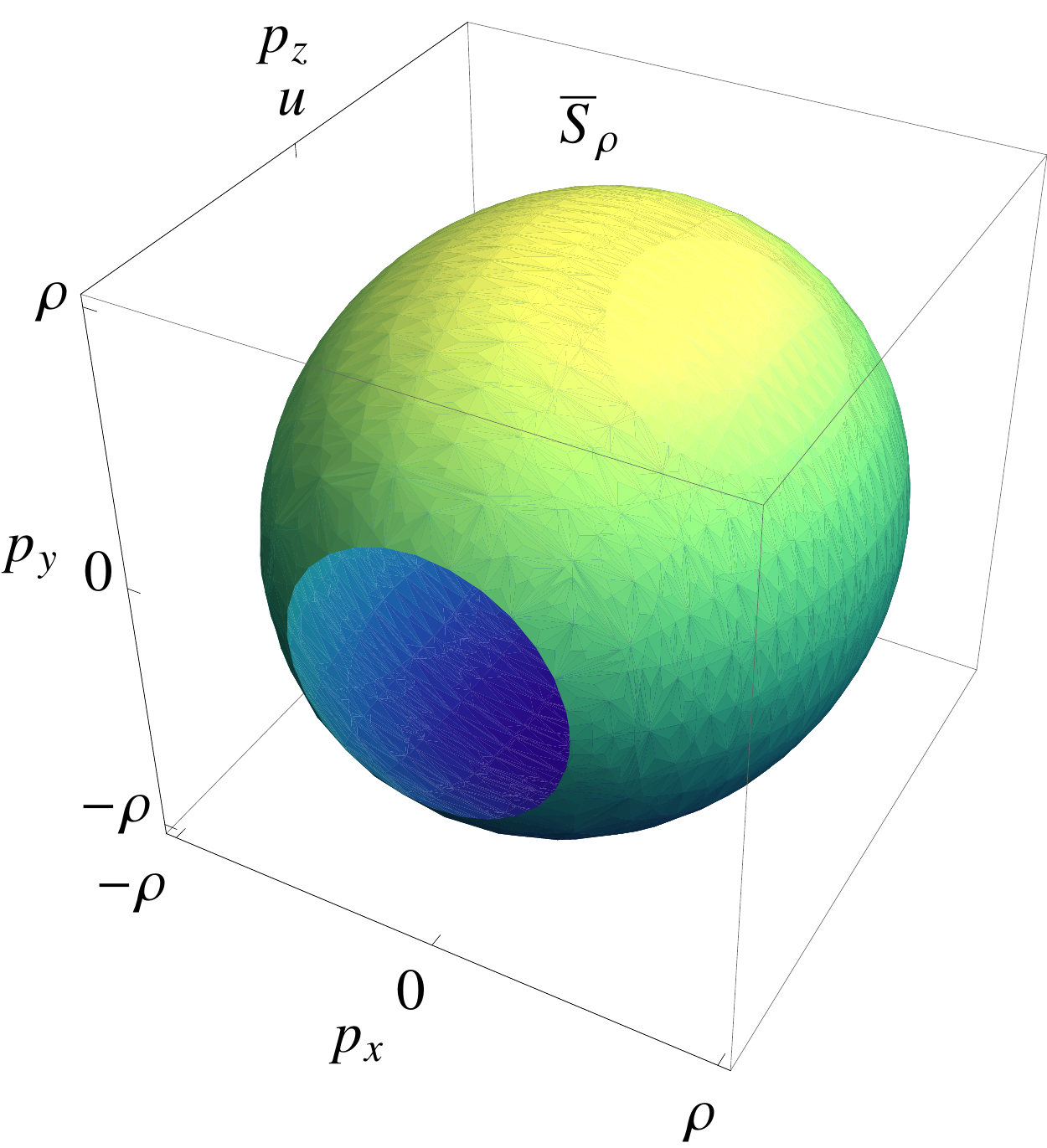}
\caption{(Color online) The pitted-olive-shaped surface $\bar{S}_\rho$ accessible for particles through diffusion along $S_\rho$.}
\label{fig:sbar}
\end{figure}

\msection{Type D} Particles can undergo nonadiabatic gyromotion [$p_\perp > \bar{p}_\perp(p_z)$] with $p_r > p_c$. Then, trapping is impossible, but multiple scatterings off the resonance at $p_x = \pm p_r$ can cause diffusion along $S_\rho$. Eventually, such trajectories can fill out the entire subset of $S_\rho$ that satisfies $p_\perp > \bar{p}_\perp(p_z)$ \cite{foot:diffusion}. This subset is further called $\bar{S}_\rho$ and has a characteristic pitted-olive shape shown in \Fig{fig:sbar}.

\section{Average current}
\label{sec:cd}

The asymmetry of $S_\rho$ along $p_z$ with respect to the origin can be used to generate an average flow, or current, parallel to $\vec{B}_0$ via surfatron acceleration of particles from an initially isotropic distribution. This is seen as follows. 

\msection{Ring distribution} First, consider an ensemble of particles initially occupying a ring $\mc{L}$ in $\vec{p}$-space, with some $(p_{\perp 0}, p_{z0} < u)$ and homogeneous distribution over $\psi$ and $\theta$, where $\theta$ is introduced via $p_{x0} = p_{\perp 0} \cos \theta$ and $p_{y0} = p_{\perp 0} \sin \theta$. If $p_{\perp 0} < \bar{p}_\perp(p_{z0})$, then $\mc{L}$ does not intersect $\Xi$, so all particles are of Type~A and will approximately retain their initial $(p_\perp, p_z)$, yielding no added current. Suppose now that $p_{\perp 0} \ge \bar{p}_\perp(p_{z0})$, so $\mc{L}$ intersects $\Xi$ at some $p_x = \pm p_r$, corresponding to $\theta_{1,2}$ [\Fig{fig:geom}(b)]. Then, trajectories of other types are realized, and current can be generated. In particular, at $p_r > p_c$, the current is produced through diffusion, for all trajectories are of Type~D and thus eventually cover the entire $\bar{S}_\rho$, changing the average velocity~by $\favr{\Delta p_z} > 0$ \cite{foot:diffusion}. In contrast, at $p_r < p_c$, the acceleration is ballistic and takes less time (of the order of $\Omega^{-1}$). Below, we assume this latter mechanism and discuss it in more detail.

\begin{figure}
\centering
\includegraphics[width=.45\textwidth]{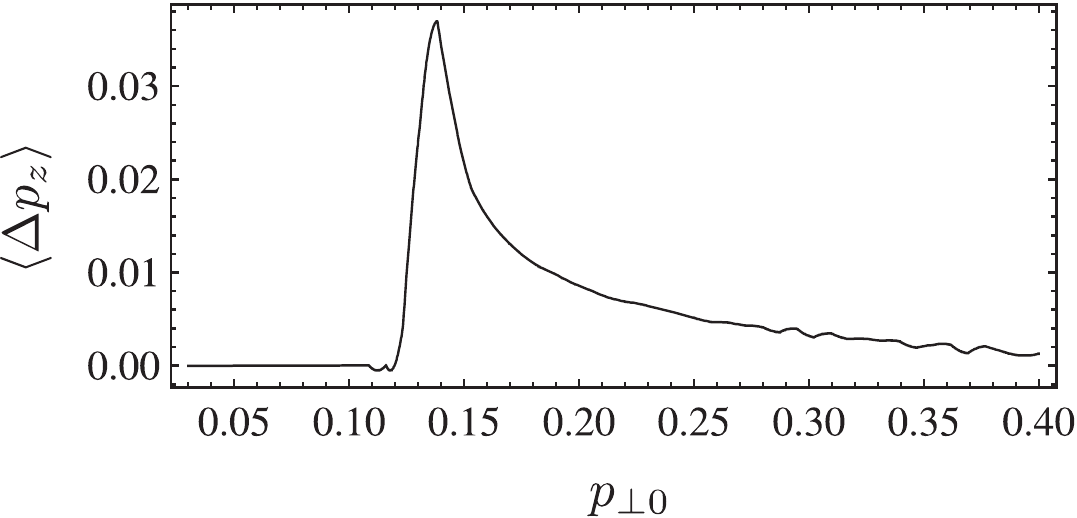}
\caption{Numerical calculation of $\favr{\Delta p_z}(p_{\perp 0})$ by tracing individual trajectories of $2^{16}$ particles (for each $p_{\perp 0}$), which were initially distributed in $\vec{p}$-space over a ring $\mc{L}$ with fixed $(p_{\perp 0}, p_{z0})$ homogeneously in $\psi$ and $\theta$. Shown is a snapshot at $t = 2\pi/\Omega$, for the same parameters as in \Fig{fig:traj}. At $p_{\perp 0} < \bar{p}(p_{z0})$, $\mc{L}$ does not intersect $\Xi$, so there is no current. At $p_{\perp 0} = \bar{p}(p_{z0})$, the current is maximal, because all trapped particles are of Type~B. At $p_{\perp 0} > \bar{p}(p_{z0})$, particles of Type~C appear, reducing $\favr{\Delta p_z}$; yet, the latter still remains positive.}
\label{fig:j}
\end{figure}

Clearly, for given $p_{z0}$, the maximum added current is generated when $p_{\perp 0} = \bar{p}_\perp(p_{z0})$, \ie when $\mc{L}$ is tangent to $\Xi$, so $\theta_1 = \theta_2$ [\Fig{fig:geom}(c)]. This is because all trapped particles are either of Type~B or remain on $\mc{L}$ forever, so $\Delta p_z \ge 0$ for \textit{each} particle in this case. At larger $p_{\perp 0}$, some of trapped particles are of Type~C. They will spend some time in the region $p_z < p_{z0}$, thus reducing the current somewhat. Still, even in this case one can show that $\favr{\Delta p_z} \ge 0$, as also seen numerically (\Fig{fig:j}).

\msection{General case} To assess the current that could be produced in plasma (\ie by particles that initially occupy a three-dimensional volume, rather than a ring, in $\vec{p}$-space), one only needs to integrate over the contributions yielded by each $\mc{L}(p_{\perp 0}, p_{z0})$. Since each $\mc{L}$ produces $\favr{\Delta p_z} > 0$, the total added current will also be nonzero. Remarkably, even particles with $p_{z0} \ll u$ will contribute here (\Fig{fig:j}). This makes surfatron acceleration very different from traditional current drive schemes \cite{ref:fisch87}, because, in the considered range ($\Omega \ll 1$), those are bound to rely on Cherenkov resonance.

Finally, notice that electromagnetic waves of the type discussed in \Ref{ref:vasiliev11} can also drive current similarly. In fact, since no particles attain negative $\Delta p_z$ in that case (assuming $p_{z0} < u$ and except in the diffusive regime), the total added current may be even larger. However, the actual amount of current that can be produced through surfatron acceleration in plasma is yet to be calculated.

\section{Conclusions}
\label{sec:conc}

In this brief note, we study surfatron acceleration of nonrelativistic particles by electrostatic waves propagating obliquely to weak dc magnetic field $\vec{B}_0$. For the particular case when the wave vector $\vec{k}$ is perpendicular to $\vec{B}_0$, the corresponding result of \Ref{ref:dawson83} is recovered. Otherwise, acceleration along magnetic field is possible, like in \Ref{ref:vasiliev11}; yet, the particle dynamics in an electrostatic wave can be qualitatively different from that in an electromagnetic wave. We also find that, in plasma, surfatron mechanism can produce an average parallel flow, or current, even when the wave frequency is much larger than the gyrofrequency and the wave phase velocity is much larger than particle initial velocities.

The work was supported by the NNSA SSAA Program through DOE Research Grant No. DE274-FG52-08NA28553 and by the U.S. DOE through Contract No. DE-AC02-09CH11466.


\begin{thebibliography}{10}
\bibitem{foot:frame}
Assumed here is the frame traveling along $z$ axis with velocity $k_z^{-1}$.
  Other coordinate transformations could also render the wave stationary but
  would change~$\vec{E}$.
\bibitem{foot:eps}
Hence, $\epsilon \equiv (k R_0)^{-1}$ in \Ref{ref:vasiliev11} is equivalent to
  our $\Theta^2$, since the characteristic Larmor radius is $R_0 = V/\Omega$.
\bibitem{foot:diffusion}
A similar effect is yielded when \Eq{eq:Theta} is violated.
\end{thebibliography}

\end{document}